\documentclass{aa}

\newcommand{\E}[1]{\mbox{$\times10^{#1}$}}
\newcommand{\err}[2]{\mbox{$^{#1}_{#2}$}}
\newcommand{\Mo}{\mbox{$M_{\protect \sun}$}}

\usepackage{graphics}

\begin{document}

\thesaurus{03     % A&A Section 3: Extragalactic astronomy
           (11.13.1;  % Galaxies: clusters: general
            12.03.3;  % Cosmology: observations
            13.25.2)} % X-rays: galaxies

\title{ASCA observations of massive medium-distant clusters of galaxies. II.}

\author{
H. Matsumoto\inst{1}
\and M. Pierre\inst{2}
\and T. G. Tsuru\inst{3}
\and D. S. Davis\inst{1}}

\offprints{H. Matsumoto}
\mail{matumoto@space.mit.edu}

\institute{
Center for Space Research, Massachusetts Institute of Technology,
77 Massachusetts Avenue, NE80, Cambridge, MA 02139-4307, U.S.A.
\and
CEA/DSM/DAPNIA, Service d'Astrophysique, F-91191 Gif sur Yvette, France
\and
Department of Physics, Faculty of Science, Kyoto University,
Sakyo-ku, Kyoto 606-8502, Japan
}

\date{Received ; accepted}

\authorrunning{H. Matsumoto et al.}
\titlerunning{ASCA observations of medium-distant clusters. II.}

\maketitle

\begin{abstract}

We have selected seven medium-distant clusters of galaxies
($z \sim 0.1 - 0.3$) for multi-wavelength observations with
the goal of investigating their dynamical state. Following
Paper~I (Pierre et al. \cite{Pierre1999}) which reported the
ASCA results about two of them, we present here the analysis
of the ASCA observations of the other five clusters;
\object{RX~J1023.8-2715} (\object{A~3444}),
\object{RX~J1031.6-2607}, \object{RX~J1050.5-0236}
(\object{A~1111}), \object{RX~J1203.2-2131}
(\object{A~1451}), and \object{RX~J1314.5-2517}.  Except for
RX~J1031.6, whose X-ray emission turned out to be dominated
by an AGN, the ASCA spectra are well fitted by a
one-temperature thin thermal plasma model. We compare the
temperature-luminosity relation of our clusters with that of
nearby ones ($z<0.1$). Two clusters, RX~J1050.5 and
RX~J1023.8, show larger luminosities than the bulk of
clusters at similar temperatures, which suggests the
presence of a cooling flow. The temperature vs.
iron-abundance relationship of our sample is consistent with
that of nearby clusters.


\end{abstract}

\section{Introduction}

Statistical studies of massive clusters of galaxies -- the
largest bound entities in the universe -- can provide
important clues for cosmology.  The most useful constraints
on cosmology are provided by quantities such as dynamics,
virialization, galaxy content, intracluster medium (ICM)
enrichment, and temperature. Specifically, it is of prime
interest to detect any sign of evolution in the cluster
properties, because this is directly influenced by the mean
density of the universe and the nature of the dark matter.
However, problems with detailed cluster studies beyond
$z\sim0.5$ are their low fluxes and the limited angular
resolving power of instruments. Therefore, we have selected
7 X-ray bright medium-distant clusters from a ROSAT All-Sky
Survey flux limited sample, which are shown in
Table~\ref{tbl:sample} along with their spectroscopic
redshifts (Pierre et al. \cite{Pierre1999} ; hereafter
Paper~I). We have performed detailed multi-wavelength
observations for them at radio, infrared, optical, and X-ray
wavelengths (Pierre et al. \cite{Pierre1994a,Pierre1994b}).
Following Paper~I which reported the ASCA results of
RX~J1131.9 (A~1300) and RX~J1325.1 (A~1732) observed during
ESA time, this paper presents the analysis of the ASCA data
of the remaining five sample clusters, acquired under ISAS
time. The complementary combined analysis of the ROSAT HRI,
radio and detailed optical spectroscopic data has been
performed by L\'emonon (\cite{Lemonon1999}) and will be
published in a forthcoming comprehensive paper.  Throughout
this paper, we assumed that $H_0$ = 50 km s$^{-1}$
Mpc$^{-1}$, and $q_0$ = 0.5. The solar abundance of iron
relative to hydrogen was taken to be 4.68\E{-5} (Anders \&
Grevesse \cite{Anders1989}). The errors in this paper are
given at the 90 \% confidence level.

\section{Observations and data reduction}

All data were obtained with two solid-state imaging
spectrometers (SIS0 and SIS1) and two gas-imaging
spectrometers (GIS2 and GIS3) at the foci of four thin-foil
X-ray mirrors (XRT) on board the ASCA satellite.
Table~\ref{tbl:ASCAlog} shows the log of the ASCA
observations.  Details concerning the instruments can be
found in Burke et al. (\cite{Burke1991}), Ohashi et
al. (\cite{Ohashi1996}), Makishima et
al. (\cite{Makishima1996}), and Serlemitsos et
al. (\cite{Serlemitsos1995}), while Tanaka et
al. (\cite{Tanaka1994}) gives a general description of
ASCA. The SIS data were obtained in the 1-CCD faint mode,
and the GIS data were obtained in the normal PH mode. Those
data were screened with the standard selection criteria:
data taken in the South Atlantic Anomaly, Earth occultation,
and regions of low geomagnetic rigidity are excluded. We
also eliminated the contamination by the bright Earth,
removed hot and flickering pixels from the SIS data, and
applied rise-time rejection to exclude particle events from
the GIS data. We further applied the ``flare-cut'' criteria
for the GIS data to exclude non X-ray background events as
many as possible (Ishisaki et
al. \cite{Ishisaki1997}). After these screenings, we
obtained effective exposure times given in Table~\ref{tbl:ASCAlog}.

\section{Analysis and Results}

\begin{figure}
\resizebox{.45\hsize}{!}{Fig.1(i)\includegraphics{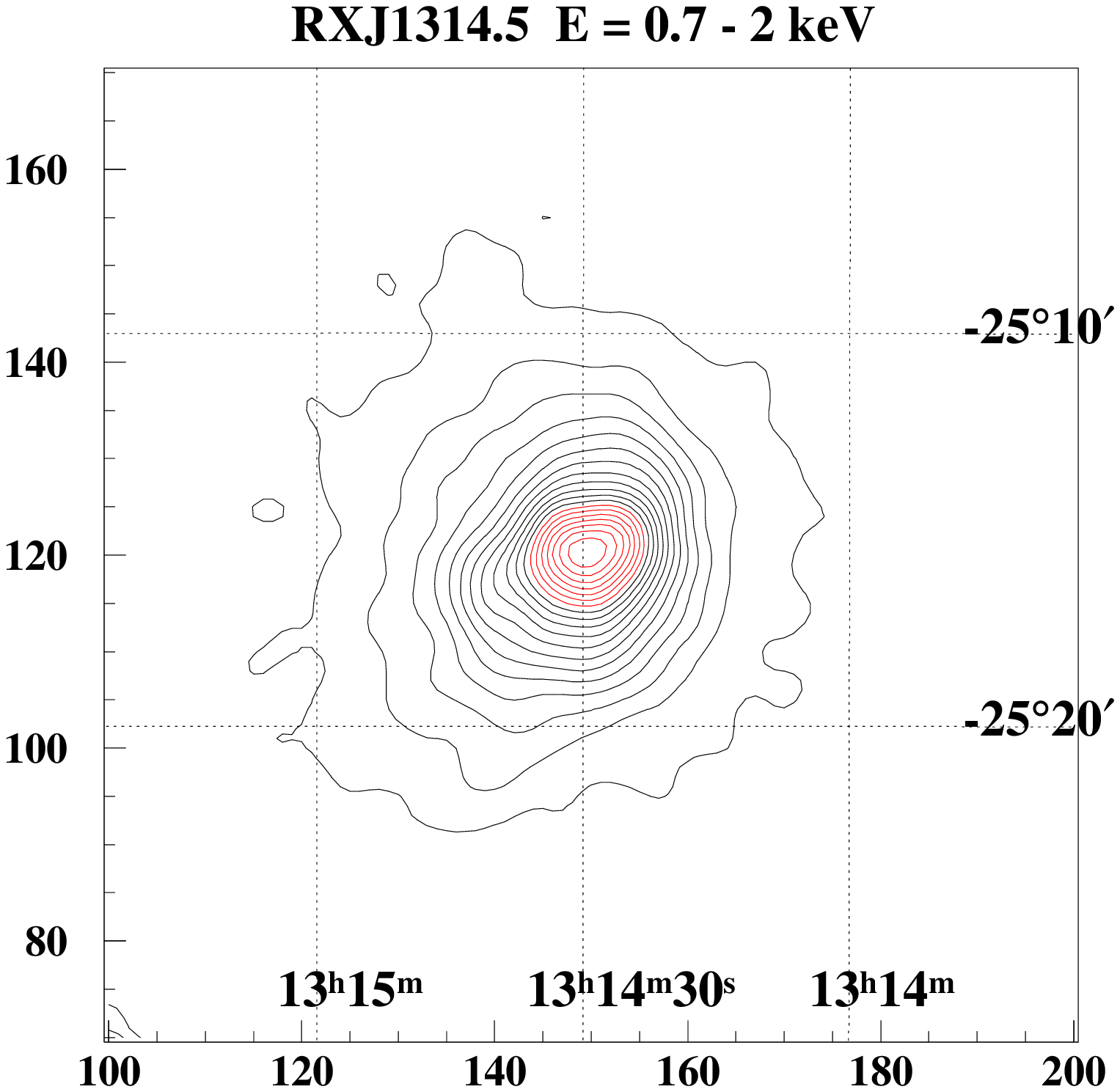}}
\resizebox{.45\hsize}{!}{Fig.1(j)\includegraphics{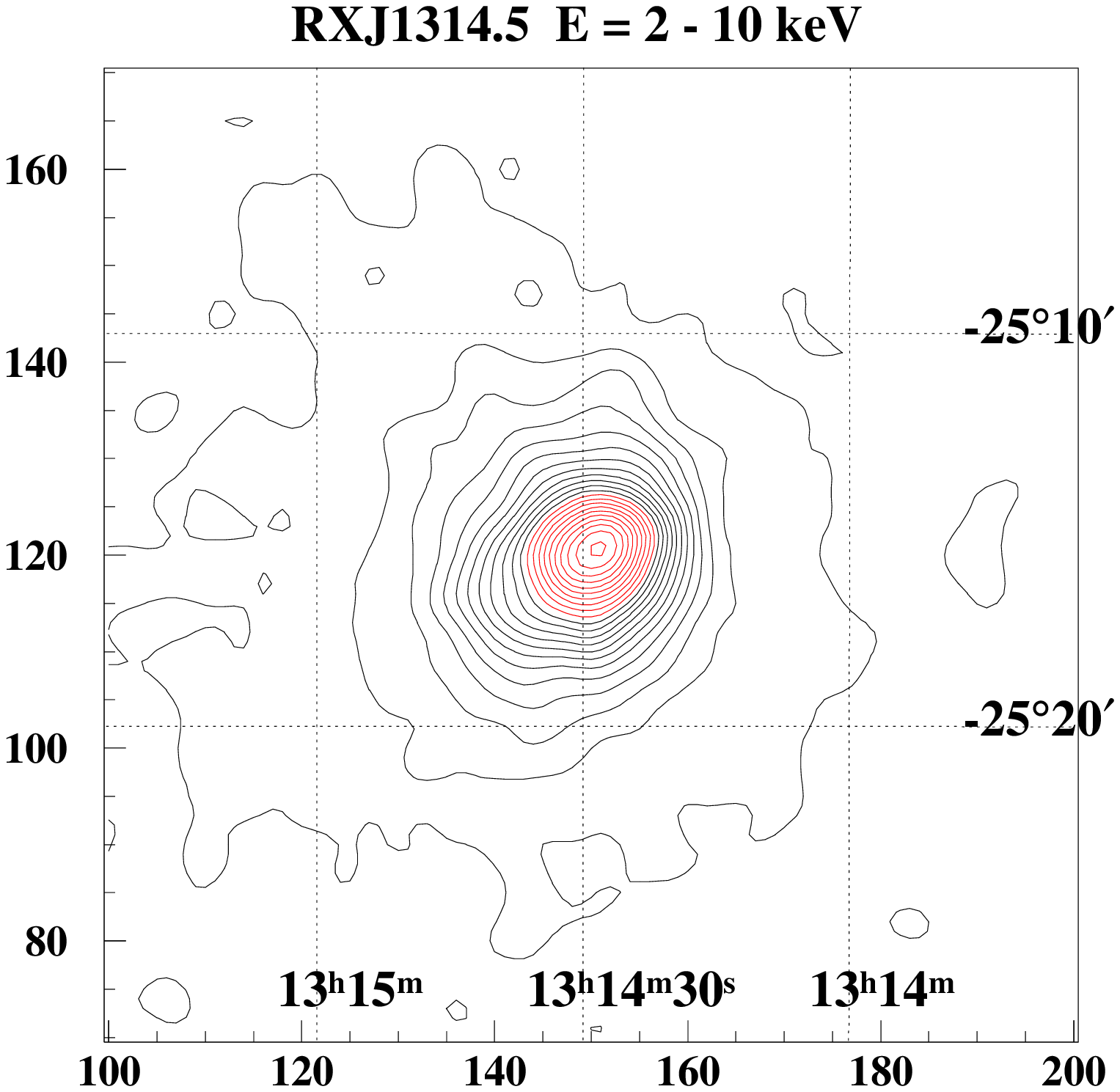}}
\caption{
ASCA GIS images of our sample clusters: 
{\bf a} RX~J1023.8 in the 0.7 -- 2 keV band, 
{\bf b} RX~J1023.8 in the 2 -- 10 keV band, 
{\bf c} RX~J1031.6 in the 0.7 -- 2 keV band, 
{\bf d} RX~J1031.6 in the 2 -- 10 keV band, 
{\bf e} RX~J1050.5 in the 0.7 -- 2 keV band, 
{\bf f} RX~J1050.5 in the 2 -- 10 keV band,
{\bf g} RX~J1203.2 in the 0.7 -- 2 keV band, 
{\bf h} RX~J1203.2 in the 2 -- 10 keV band, 
{\bf i} RX~J1314.5 in the 0.7 -- 2 keV band, 
{\bf j} RX~J1314.5 in the 2 -- 10 keV band. The pixel
size is $0.25'$ and the images have been filtered by a
Gaussian having a $\sigma$ of 2 pixels. Contour levels are
1.0, 2.0, 3.0... counts/pixel.
}
\label{fig:GISimg}
\end{figure}

From the screened data, we extracted the SIS and GIS images
of our sample clusters. We show the GIS images in the 0.7 --
2.0 keV band and in the 2.0 -- 10 keV band in
Fig.~\ref{fig:GISimg}.  RX~J1023.8, RX~J1203.2, and
RX~J1314.5 were observed twice.  However, since we found no
differences between the two observations, only the images
obtained by the first observation (i.e. the observations
labeled \#1 in Table~\ref{tbl:ASCAlog}) are shown. We found
no remarkable structures such as distortions or
substructures in the images except for RX~J1050.5: a clump
is conspicuous in the soft-band image, some $6'$ south off
the cluster center. We identify this as a foreground star
and it is described further in the Appendix.

We extracted the SIS and GIS spectra of each cluster from
the screened data. The extraction regions were circular, and
centered on each cluster. The extraction radii were $6'$ for
the GIS and $4'$ for the SIS for all clusters except
RX~J1050.5. For the data of RX~J1050.5, the extraction radii
for the GIS spectra were also restricted to $4'$ in order to
avoid contamination by the foreground star. The spectra were
then rebinned to contain at least 40 counts in each spectral
bin to utilize the $\chi^2$ technique.

Background spectra for the GIS were extracted from the blank
sky data taken during the Large Sky Survey project
(e.g. Ueda et al. \cite{Ueda1998}) with the same
data-reduction method as for the cluster data. We obtained
the SIS background spectra from a source-free region around
each cluster. We confirmed that our results did not change
significantly when using the GIS spectra taken from the
source free regions as the GIS backgrounds. In our spectral
analysis, we fitted the SIS0, SIS1, GIS2, and GIS3 spectra
of each cluster simultaneously using the XSPEC spectral
fitting package ver 10.0 (Arnaud \cite{Arnaud1996}).

\subsection{RX~J1023.8, RX~J1050.5, RX~J1203.2, and RX~J1314.5}

The extracted spectra for RX~J1023.8, RX~J1050.5,
RX~J1203.2, and RX~J1314.5 in the 3 -- 10 keV band are
fitted with a thermal bremsstrahlung model. Next, we added a
Gaussian line model to the bremsstrahlung model, and
compared the $\chi^2$ values to examine the existence of
iron K lines and to investigate whether the
redshift of the iron K lines is consistent with the redshift
in Table~\ref{tbl:sample}.  This analysis showed that the
four clusters have statistically significant Gaussian
lines. The best-fit values are shown in
Table~\ref{tbl:gauss}. Assuming the redshifts in
Table~\ref{tbl:sample}, we confirmed that the center energy
of the Gaussian lines at the rest frames of the clusters are
consistent with the iron K $\alpha$ lines from thin thermal
plasma by using the MEKAL code (Mewe et
al. \cite{Mewe1995}).

\begin{figure}
\rotatebox{-90}{
\resizebox{.45\hsize}{!}{Fig.2(c)\includegraphics{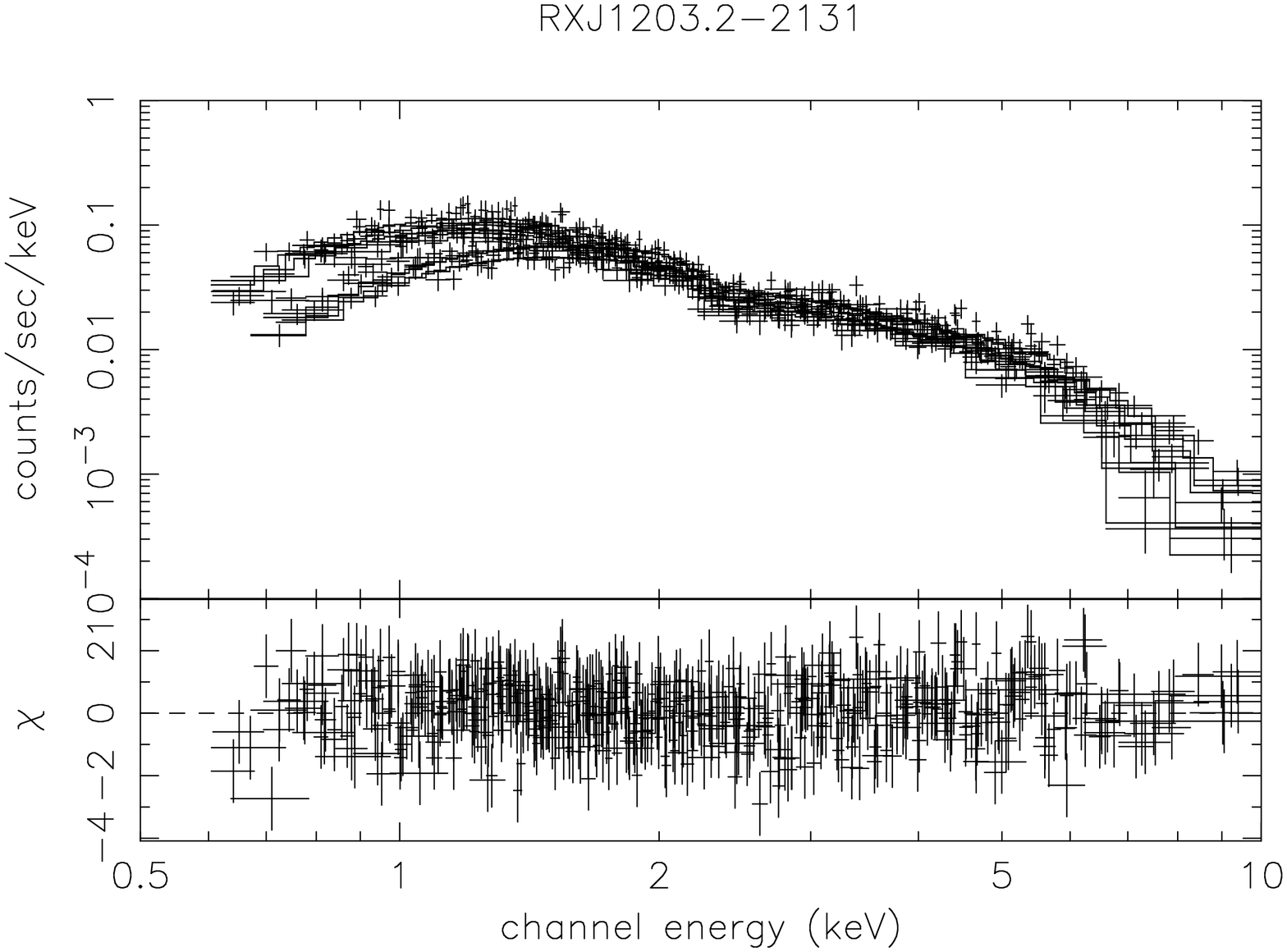}}
}
\rotatebox{-90}{
\resizebox{.45\hsize}{!}{Fig.2(d)\includegraphics{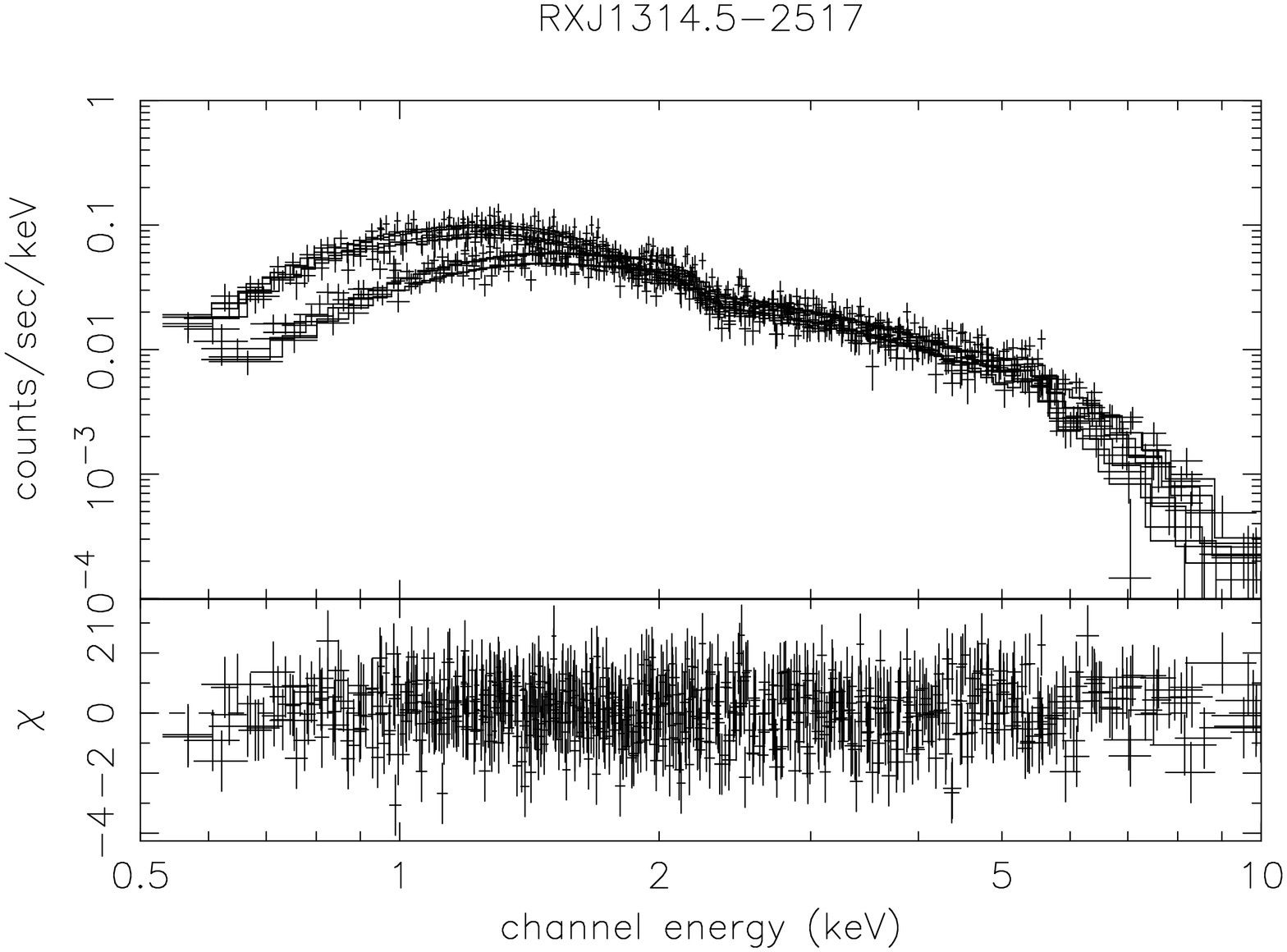}}
}
\caption{ ASCA spectra of individual clusters with the
best-fitting MEKAL model: {\bf a} RX~J1023.8, {\bf b}
RX~J1050.5, {\bf c} RX~J1203.2, {\bf d} RX~J1314.5. The
upper panel shows the SIS0, SIS1, GIS2 and GIS3 spectra
together with the best-fitting model. The lower panel shows
the residual between the data and the best-fit model.  }
\label{fig:spec}
\end{figure}

We then performed spectral fitting in the 0.6 -- 10 keV band
for the SIS, and in the 0.7 -- 10 keV band for the GIS.  We
fitted the spectra with the thin thermal plasma model (the
MEKAL model) modified by interstellar absorption.  We fixed
the redshift of each cluster to the values presented in
Table~\ref{tbl:sample}. The free parameters were the
absorption column density ($N_{\rm H}$), temperature ($kT$),
metal abundance, and normalization. Since the metal
abundance was determined mainly by the iron K $\alpha$
lines, it can be regarded as the iron abundance ($A_{\rm
Fe}$). Fig.~\ref{fig:spec} shows the spectra with the
best-fit model, and Table~\ref{tbl:mekal} shows the best-fit
parameters. The temperatures are consistent with those in
Table~\ref{tbl:gauss}, but have much smaller errors. This is
the result of using wider energy ranges which improved the
data statistics. The $N_{\rm H}$ values obtained here are
consistent with the Galactic values (Dickey \& Lockman
\cite{Dickey1990}) except for RX~J1023.8. The Galactic
$N_{\rm H}$ toward RX~J1023.8 is 5.4\E{20} cm$^{-2}$. The
excess column density of 7.9\E{20} cm$^{-2}$ suggests that
RX~J1023.8 has a cooling flow (White et
al. \cite{White1991}).

\subsection{RX~J1031.6}

L\'emonon (\cite{Lemonon1999}) analyzed the radial profile
of RX~J1031.6 obtained with the ROSAT HRI and found that the
profile is consistent with a point source which coincides
with an optical galaxy at a redshift of 0.2441,
which has been measured during a spectroscopic run at the
ESO 3.6m telescope (Pierre et al \cite{Pierre1994a}). This
suggests that the X-ray emission is dominated by an AGN, and
consequently, that the object was misclassified as a
``bright" cluster from the low resolution RASS data. From
the optical point of view, this cluster indeed appears as a
loose group.  The radial profiles of the SIS and GIS images
are also consistent with a point source.

We fitted the spectra with the MEKAL model modified by the
absorption. The redshift is fixed to the value in
Table~\ref{tbl:sample} ($z$ = 0.247). The best-fit
parameters were $N_{\rm H}$ = 1.0\err{+1.3}{-1.0}\E{20}
cm$^{-2}$, $kT$ = 5.0\err{+0.3}{-0.2} keV, $A_{\rm Fe}$ =
0.07\err{+0.04}{-0.05} solar, and $\chi^2/d.o.f.$ =
616.89/575.  Though the $\chi^2$ value suggests that the
fitting is acceptable, we found residual structures between
the data and the best-fit model above 5 keV.

\begin{figure}
\rotatebox{-90}{
\resizebox{.5\hsize}{!}{\includegraphics{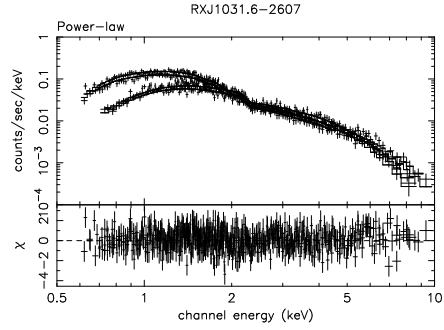}}
}
\caption{ASCA spectrum of RX~J1031.6 with the best-fitting power-law model.}
\label{fig:RXJ1031spec}
\end{figure}

As a next step, we tried a power-law model modified by the
absorption, since X-ray spectra from typical AGNs can be
roughly described by the power-law model (e.g. Mushotzky et
al. \cite{Mushotzky1993}). The free parameters were the
column density, photon index ($\Gamma$), and normalization.
The best-fit parameters were $N_{\rm H}$ =
14.1\err{+1.7}{-1.6}\E{20} cm$^{-2}$, $\Gamma$ =
2.1\err{+0.1}{-0.1}, and $\chi^2/d.o.f.$ = 519.54/576.  In
this case, the $\chi^2$ value is much lower than with the
MEKAL model fitting. Therefore, it is reasonable to conclude
that the X-ray emission from RX~J1031.6 is dominated by an
AGN, though the best-fit photon index is a little higher
than the typical value ($\sim$ 1.7). The spectra of
RX~J1031.6 are shown in Fig.~\ref{fig:RXJ1031spec} along
with the best-fitting power-law model. The flux and
unabsorbed luminosity obtained with the GIS in the 0.5 -- 10
keV band are 7.6\E{-12}~erg~s$^{-1}$~cm$^{-2}$, and
2.7\E{45}~erg~s$^{-1}$, respectively, for a redshift of
0.244, which is the redshift of the central galaxy.  The
$N_{\rm H}$ value obtained here is much larger than the
Galactic value (5.3\E{20} cm$^{-2}$; Dickey \& Lockman
\cite{Dickey1990}), and this means that the X-ray emission
from the AGN suffers absorption due to its host galaxy.

There are two ROSAT HRI archival data of RX~J1031.6.  Each
of them was observed on May 25, 1996, and Dec. 19, 1996.  We
analyzed them and found that the HRI counting rate of
RX~J1031.6 was ($4.0\pm0.5$)\E{-2}~c~s$^{-1}$ on May 25,
1996, and ($2.4\pm0.4$)\E{-2}~c~s$^{-1}$ on Dec. 19,
1999. Assuming the best-fitting power-law model to the ASCA
data, these counting rates correspond to the fluxes of
($2.8\pm0.4$)\E{-12}~erg~s$^{-1}$~cm$^{-2}$ and
($1.7\pm0.3$)\E{-12}~erg~s$^{-1}$~cm$^{-2}$ in the 0.5 --
10~keV band. Thus, we found that the AGN has time
variability.

\subsection{Cooling flow clusters (RX~J1023.8 and RX~J1050.5)}

L\'emonon (\cite{Lemonon1999}) analyzed the ROSAT HRI images
of RX~J1050.5 and RX~J 1023.8, and found that the radial
profiles were strongly peaked at the cluster centers. This
suggests they have a cooling flow. Considering the cooling
time at the cluster center, the mass deposition rates were
found to be of the order of 1400 \Mo~yr$^{-1}$ for
RX~J1050.5 and 2500 \Mo~yr$^{-1}$ for RX~J1023.8 (L\'emonon
\cite{Lemonon1999}). The excess column density above the
Galactic value found in RX~J1023.8 also supports the
presence of the cooling flow in RX~J1023.8, though the
column density of RX~J1050.5 is marginally consistent with
the Galactic value of 4.1\E{20}~cm$^{-2}$ (Dickey \& Lockman
\cite{Dickey1990}).

Therefore, we tried to fit the spectra of the two clusters
with a two-temperature MEKAL model modified by the
absorption, which can be represented by $N_{\rm
H}\times({\rm MEKAL(cool)} + {\rm MEKAL(hot)})$.  We fixed
the redshifts to the values in Table~\ref{tbl:sample}.  We
also assumed the metal abundance of the hot component is the
same as that of the cool component. Then, the free
parameters were the column density, the temperatures of the
hot and cool components, the metal abundance, and the
normalizations of the cool and hot components. However, we
obtained no significant improvement compared with the
one-temperature MEKAL model.  Thus, it appears not possible
to spectroscopically assess the multi-temperature structure
of the ICM with the current data statistics (the number of
X-ray photons used in the analysis is 9080 for RX~J1050.5
and 9444 for RX~J1023.8).

\section{Discussion}

\begin{figure}
\rotatebox{-90}{
\resizebox{.5\hsize}{!}{\includegraphics{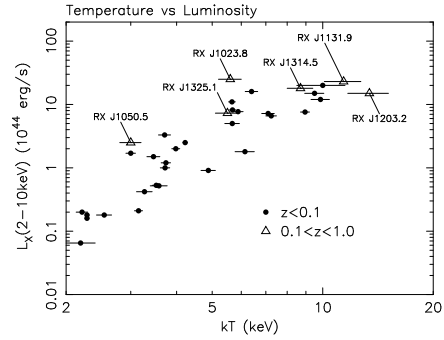}}
}
\caption{ Temperature-luminosity relation. Triangles show
our results including Paper I (Pierre et
al. \cite{Pierre1999}) except for RX~J1031.6-2607, and dots
show nearby clusters ($z<0.1$) (Matsumoto et
al. \cite{Matsumoto2000}).}
\label{fig:kT_Lx}
\end{figure}

The temperature-luminosity relation of our sample clusters
excluding RX~J1031.6, plus the two clusters (RX~J1131.9 and
RX~J1325.1) presented in Paper~I is shown in
Fig.~\ref{fig:kT_Lx}. In Fig.~\ref{fig:kT_Lx}, we also
display the data for nearby clusters ($z<0.1$) whose
temperatures and luminosities were determined by ASCA
(Matsumoto et al. \cite{Matsumoto2000}). These two samples
are reduced in the same manner so that cross-calibration
variations between different instruments should not be an
issue.  We notice that the cooling flow clusters, RX~J1023.8
and RX~J1050.5, have rather large luminosities in comparison
to the other clusters of similar temperatures. This is an
indicator in favor of the presence of large amounts of cold
gas (e.g. Fabian \cite{Fabian1994}).

The other clusters in our sample are in good agreement with
the nearby cluster relationship, which is consistent with
previous claims that there is no evolution in the
temperature-luminosity relation at $z<1.0$ (Mushotzky \&
Scharf \cite{Mushotzky1997b}; Matsumoto et
al. \cite{Matsumoto2000}).  The ROSAT HRI images of
RX~J1203.2 and RX~J1314.5 show signs of structure, and they
are thought to be merging systems (L\'emonon
\cite{Lemonon1999}). However, the clusters do not depart
from the average temperature-luminosity correlation.  
The hottest cluster, RX~J1203.2, may suggest that the
temperature-luminosity relation flattens at high
temperatures.  However, this flattening is partly due to the
definition of our luminosity, because we use the luminosity
in the 0.5 -- 10~keV band and the temperature of the cluster
is outside of this energy band. We also should note that the
temperature-luminsoty relation has significant
dispersion. In fact, a high temperature cluster MS~1054-0321
($kT=12.3^{+3.1}_{-2.2}$~keV) at $z=0.829$ does not show the
evidence of the flattening (Donahue et al. \cite{Donahue1998}).

\begin{figure}
\rotatebox{-90}{
\resizebox{.5\hsize}{!}{\includegraphics{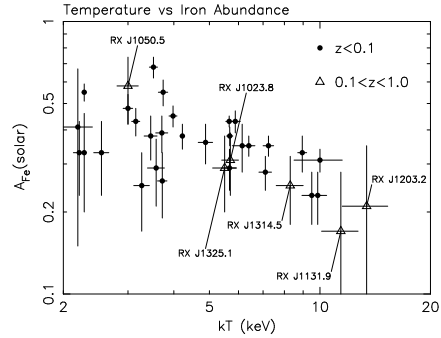}}
}
\caption{ 
Temperature-iron abundance relation. The
symbols are the same as in Fig \ref{fig:kT_Lx}.}
\label{fig:kT_Z}
\end{figure}

The temperature-iron abundance relation is seen
Fig.~\ref{fig:kT_Z}.  There is no clear difference between
our sample clusters and the nearby ones, which confirms the
previous result that there is no evolution in the iron
abundance at $z<1.0$ (Mushotzky \& Loewenstein
\cite{Mushotzky1997a}; Matsumoto et
al. \cite{Matsumoto2000}).  

\begin{figure}
\rotatebox{-90}{
\resizebox{.5\hsize}{!}{\includegraphics{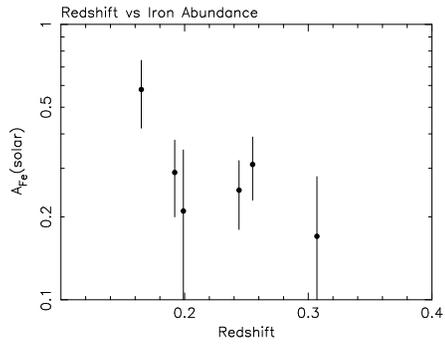}}
}
\caption{ 
Redshift-iron abundance relation of our sample clusters.}
\label{fig:red_metal}
\end{figure}

We also investigated the redshift-iron abundance relation
(Fig.~\ref{fig:red_metal}). As already noted by Mushotzky \&
Loewenstein \cite{Mushotzky1997a}, we can see no evidence
for the evolution of the iron abundance with redshift.  The
mean abundance of our sample is $0.30\pm0.05$ solar.

\section{Conclusion}

We have analyzed the ASCA data of five medium-distant
clusters of galaxies; RX~J1023.8, RX~J1031.6, RX~J1050.5,
RX~J1203.2, and RX~J1314.5. Except for RX~J1031.6 whose
X-ray emission is dominated by an AGN, we were able to fit
the ASCA spectra of the clusters with the one-temperature
thermal plasma model. We compared the temperature-luminosity
relation of our sample clusters with that of the nearby
clusters ($z<0.1$). We found that RX~J1023.8 and RX~J1050.5
have rather large luminosities in the temperature-luminosity
plane. This can be explained by the presence of a cooling
flow, although the present statistics do not allow
multi-temperature fitting. The other clusters lay well
within the mean temperature-luminosity relation defined by
the low-redshift clusters.  In addition, the metallicity of
our sample clusters are in good agreement with the local
temperature-iron abundance relationship, which is consistent
with the previous findings out to $z<1.0$.

\begin{acknowledgements} 
We would like to thank the ASCA team members for their
support. HM is supported by the JSPS Postdoctoral
Fellowships for Research Abroad.  This research has made use
of the SIMBAD database, operated at CDS, Strasbourg, France.
\end{acknowledgements}

\clearpage

%table 1
\begin{table}
\caption{ASCA cluster sample.}
\label{tbl:sample}
\begin{tabular}{lll} \hline \hline
ROSAT ID	&Abell ID	&redshift \\\hline
RX~J1023.8-2715	&Abell 3444	&0.255\\
RX~J1031.6-2607	&		&0.247\\
RX~J1050.5-0236	&Abell 1111	&0.165\\
RX~J1131.9-1955	&Abell 1300	&0.307\\
RX~J1203.2-2131	&Abell 1451	&0.199\\
RX~J1314.5-2517	&		&0.244\\
RX~J1325.1-2013	&Abell 1732	&0.192\\ \hline
\end{tabular}
\end{table}

%table 2
\begin{table}
\caption{ASCA observation log of the 5 clusters.}
\label{tbl:ASCAlog}
\begin{tabular}{lllllll} \hline \hline
\multicolumn{2}{l}{Name}	&Date (UT)	&\multicolumn{4}{c}{Exposure (s)$^{\dag}$}\\ \cline{4-7}
		&	&	&SIS0	&SIS1	&GIS2	&GIS3\\ \hline
RX~J1023.8 	&\#1	&Jun 5, 1997	&19327	&19157	&20266	&20266\\
 		&\#2	&May 21, 1998	&6121	&6149	&6174	&6174\\
RX~J1031.6	&	&Dec 27, 1995	&46305	&46333	&54287	&53695\\
RX~J1050.5	&	&Dec 18, 1995	&28115	&27774	&28807	&28801\\
RX~J1203.2	&\#1	&Jun 17, 1996	&20114	&20038	&22952	&22944\\
		&\#2	&Dec 25, 1996	&13681	&13500	&14026	&14026\\
RX~J1314.5	&\#1	&Jan 31, 1996	&27642	&27747	&29234	&29244\\
		&\#2	&Feb 3, 1996	&27183	&26888	&28905	&28895\\ \hline
\end{tabular}\\
$\dag$ Total exposure time after the data screening described in section 2.
\end{table}

%table 3
\begin{table}[t]
\caption{Results of the bremsstrahlung plus Gaussian model fitting.}
\label{tbl:gauss}
\begin{tabular}{llllll}\hline\hline
Name		&$kT^a$			&$E_{\rm c}^b$		&$\sigma^c$		&$EW^d$	&$\chi^2/d.o.f.$\\
		&(keV)			&(keV)			&(eV)		&(eV)	\\ \hline
RX~J1023.8	&5.7\err{+1.1}{-0.8}	&5.3\err{+0.1}{-0.1}	&77\err{+104}{-77}	&290	&131.51/118\\
RX~J1050.5	&3.0\err{+1.1}{-1.0}	&5.7\err{+0.3}{-0.1}	&130\err{+620}{-130}	&657	&29.59/32\\
RX~J1203.2	&12.7\err{+4.2}{-2.6}	&5.4\err{+0.3}{-0.1}	&1.4\err{+339}{-1.4}	&137	&148.77/159\\
RX~J1314.5	&9.6\err{+2.1}{-1.5}	&5.4\err{+0.1}{-0.2}	&0\err{+240}{\cdots}	&140	&199.79/208\\ \hline
\end{tabular}\\
The errors are described at the 90\% confidence limits for one parameter.\\
a: The temperature of the bremsstrahlung model.\\
b: The center energy of the Gaussian line model.\\
c: The sigma of the Gaussian line model.\\
d: The equivalent width of the Gaussian line model.\\
\end{table}

%table 4

\begin{table}[t]
\caption{Results of the MEKAL model fitting.}
\label{tbl:mekal}
\begin{tabular}{lllllll}\hline\hline
Name		&$N_{\rm H}$		&$kT$			&$A_{\rm Fe}$		&$F_{\rm X}^{\dag}$	&$L_{\rm X}^{\S}$	&$\chi^2/d.o.f.$	\\
	&($10^{20}$cm$^2$)		&(keV)			&(solar)		&($10^{-12}$ erg/s/cm$^2$)	&($10^{45}$ erg/s)	\\ \hline
RX~J1023.8	&13.3\err{+1.8}{-1.8}	&5.6\err{+0.4}{-0.3}	&0.30\err{+0.08}{-0.07}	&11 (7.4)	&4.0 (2.5)	&456.30/489\\
RX~J1050.5	&6.6\err{+2.7}{-2.5}	&3.0\err{+0.2}{-0.2}	&0.55\err{+0.14}{-0.13}	&3.5 (1.8)	&0.49 (0.25)	&196.93/209\\
RX~J1203.2	&5.2\err{+1.7}{-1.7}	&13.4\err{+1.9}{-1.5}	&0.24\err{+0.13}{-0.15}	&12 (8.5)	&2.1 (1.5)	&484.09/534\\
RX~J1314.5	&8.7\err{+1.7}{-1.6}	&8.7\err{+0.7}{-0.6}	&0.25\err{+0.07}{-0.08}	&9.1 (6.4)	&2.7 (1.8)	&664.13/729\\ \hline
\end{tabular}\\
The errors are described at the 90\% confidence limits for one parameter.\\
$\dag$: The GIS flux in the 0.5 -- 10 keV band. The values in the brackets are in the 2 -- 10 keV band.\\
$\S$: The GIS luminosity in the 0.5 -- 10 keV band. These values are absorption
corrected. The values in the brackets are in the 2 -- 10 keV band.
\end{table}

\clearpage

%Fig. 1

\begin{figure}
\resizebox{.45\hsize}{!}{Fig.1(a)\includegraphics{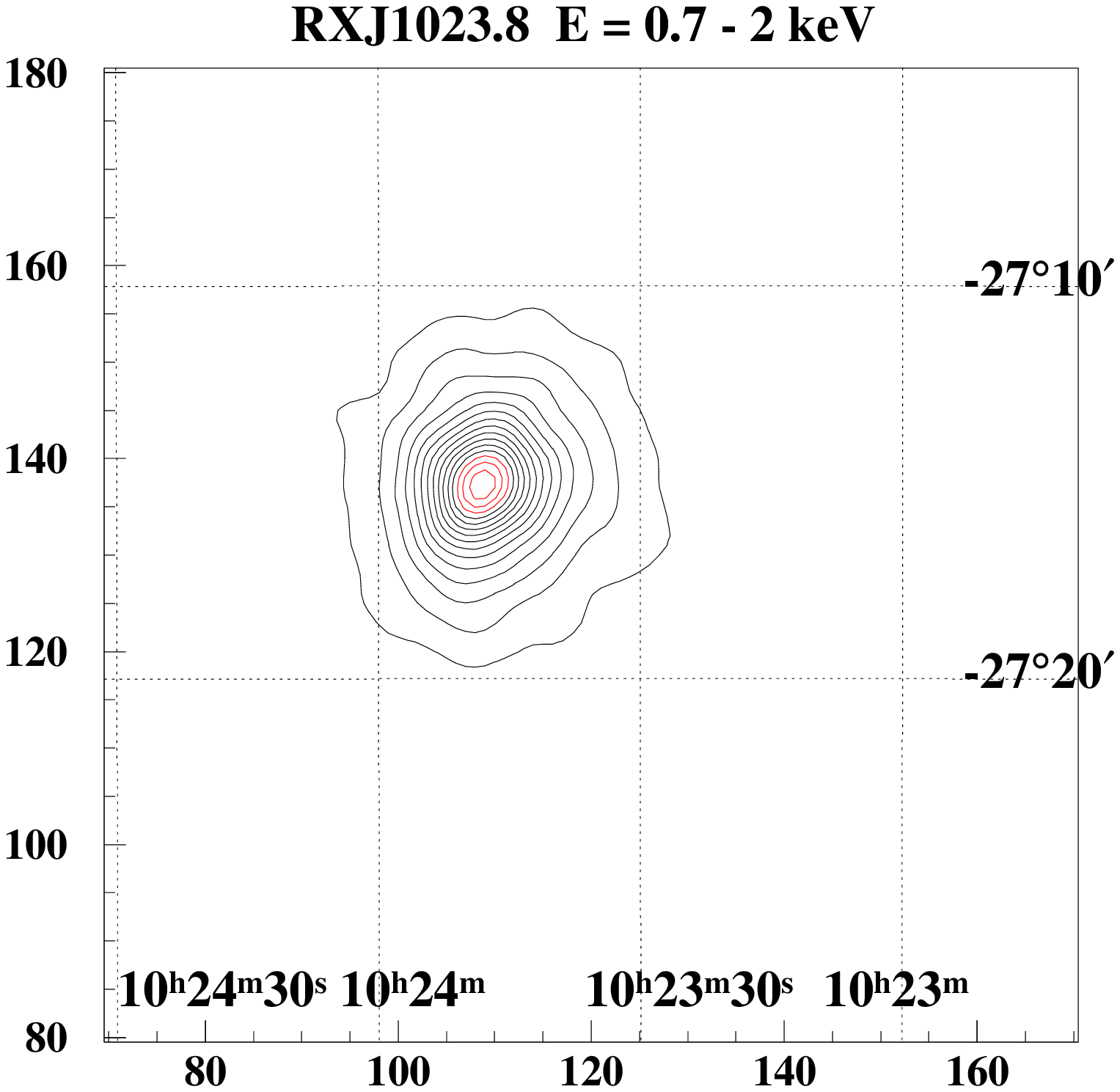}}
\resizebox{.45\hsize}{!}{Fig.1(b)\includegraphics{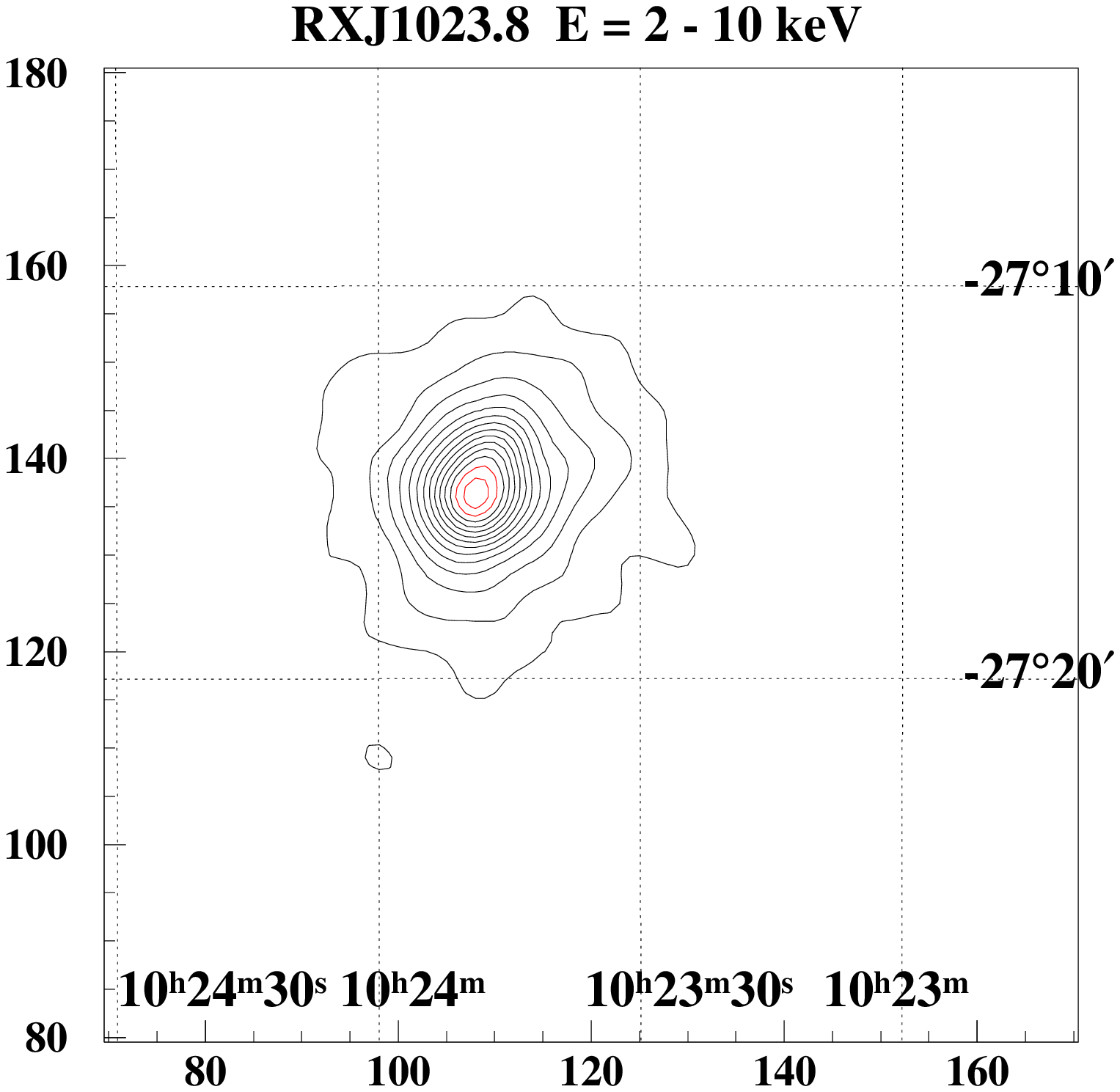}}
\end{figure}

\begin{figure}
\resizebox{.45\hsize}{!}{Fig.1(c)\includegraphics{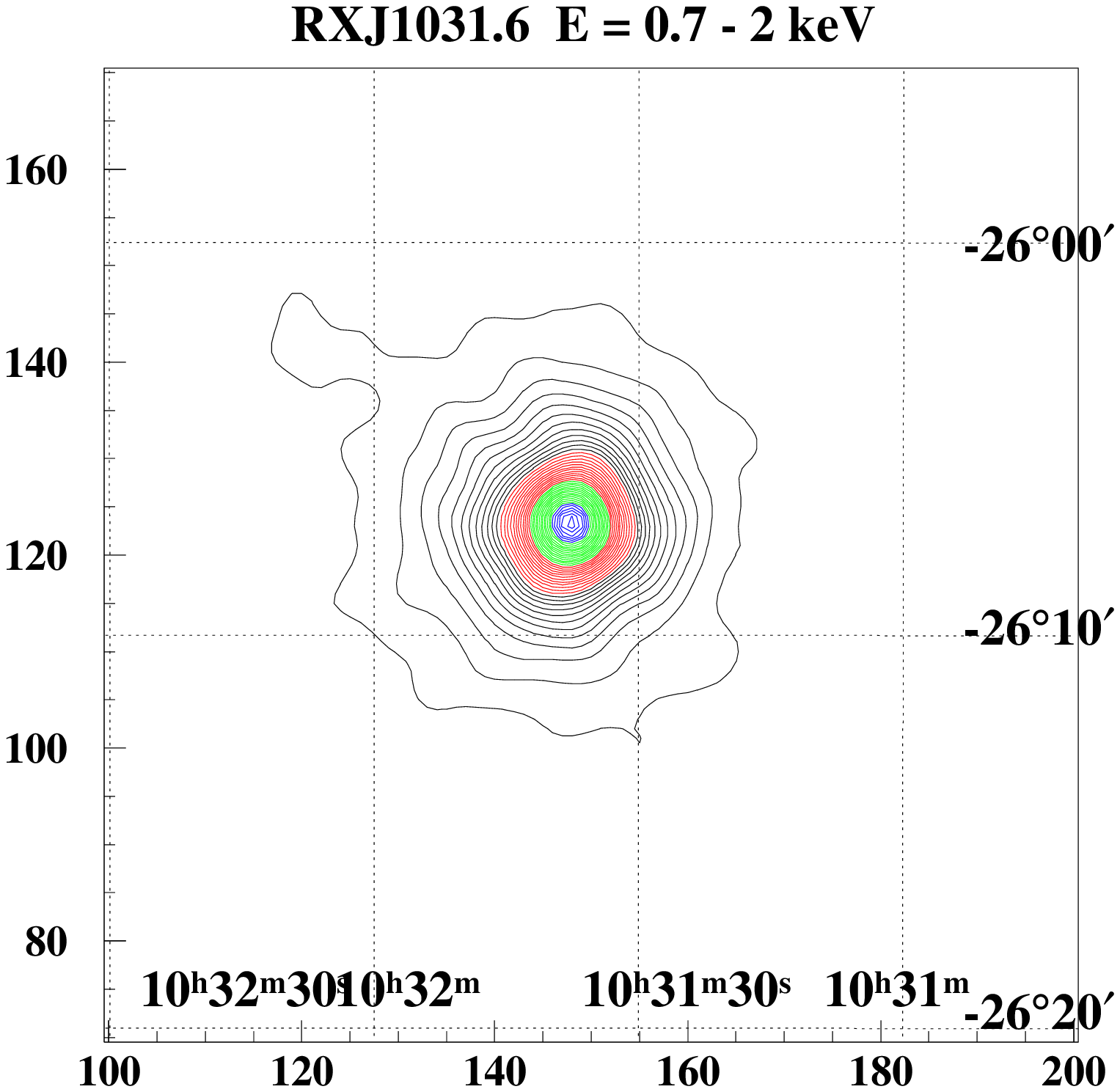}}
\resizebox{.45\hsize}{!}{Fig.1(d)\includegraphics{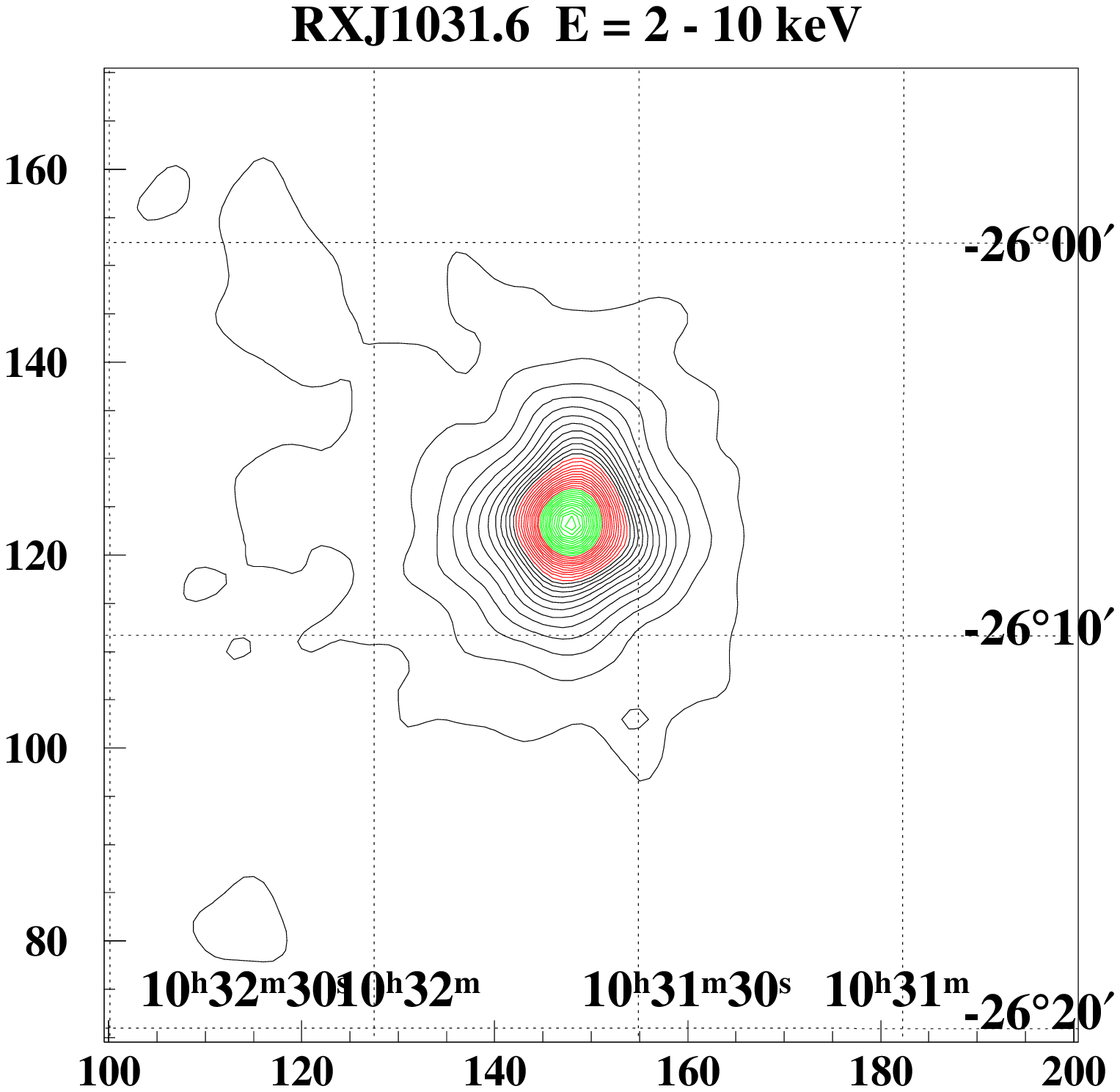}}
\end{figure}

\begin{figure}
\resizebox{.45\hsize}{!}{Fig.1(e)\includegraphics{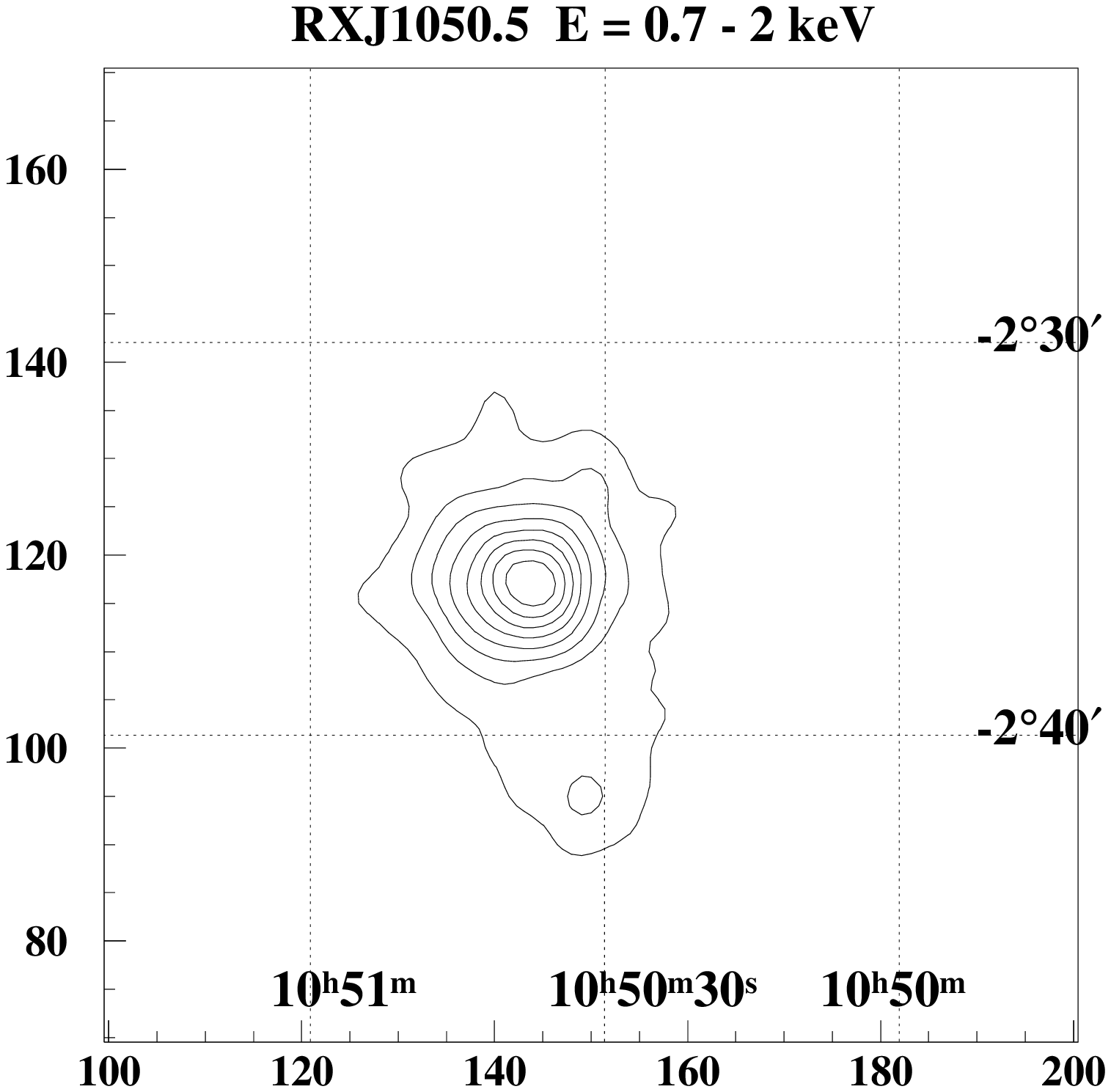}}
\resizebox{.45\hsize}{!}{Fig.1(f)\includegraphics{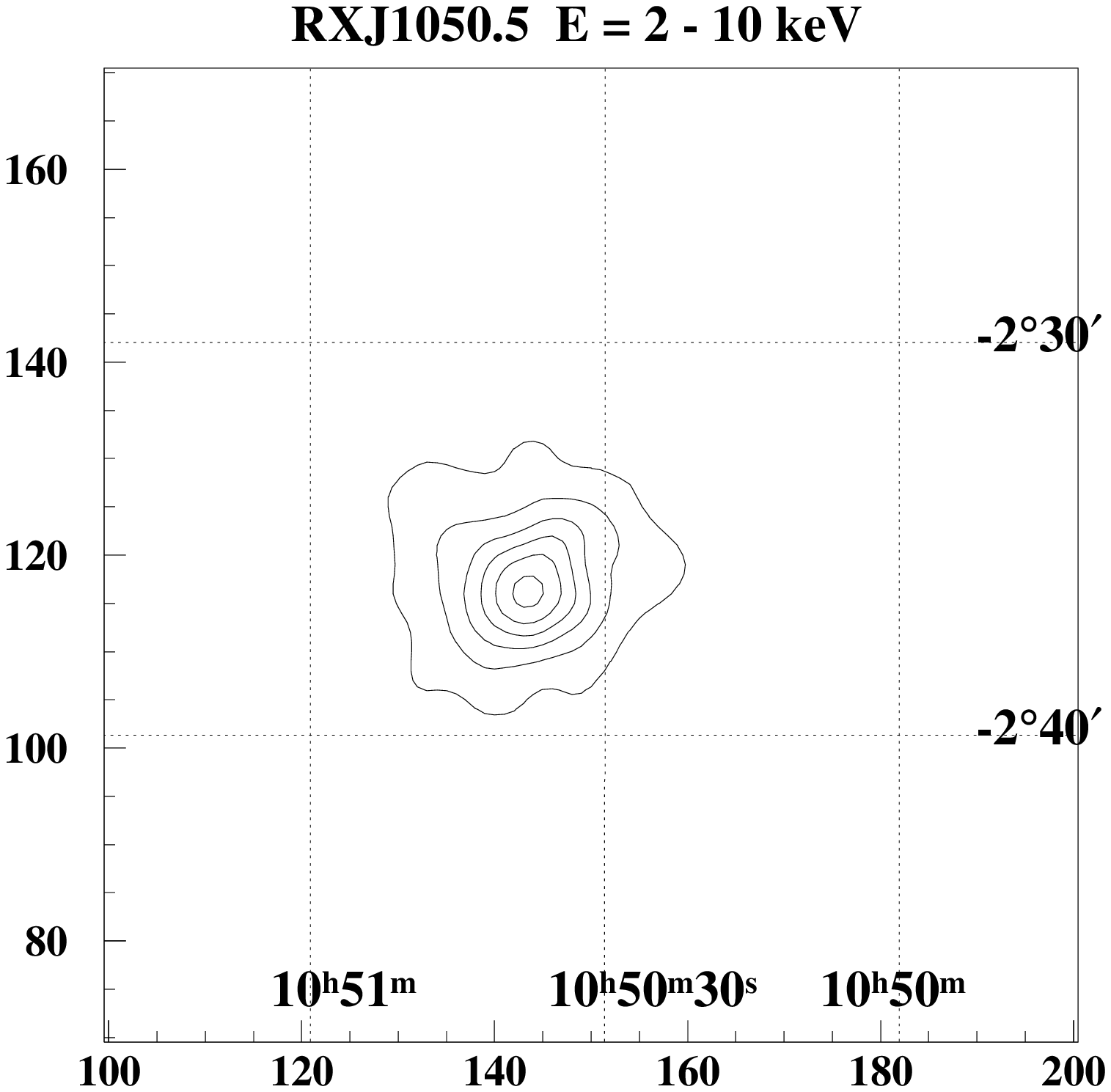}}
\end{figure}

\begin{figure}
\resizebox{.45\hsize}{!}{Fig.1(g)\includegraphics{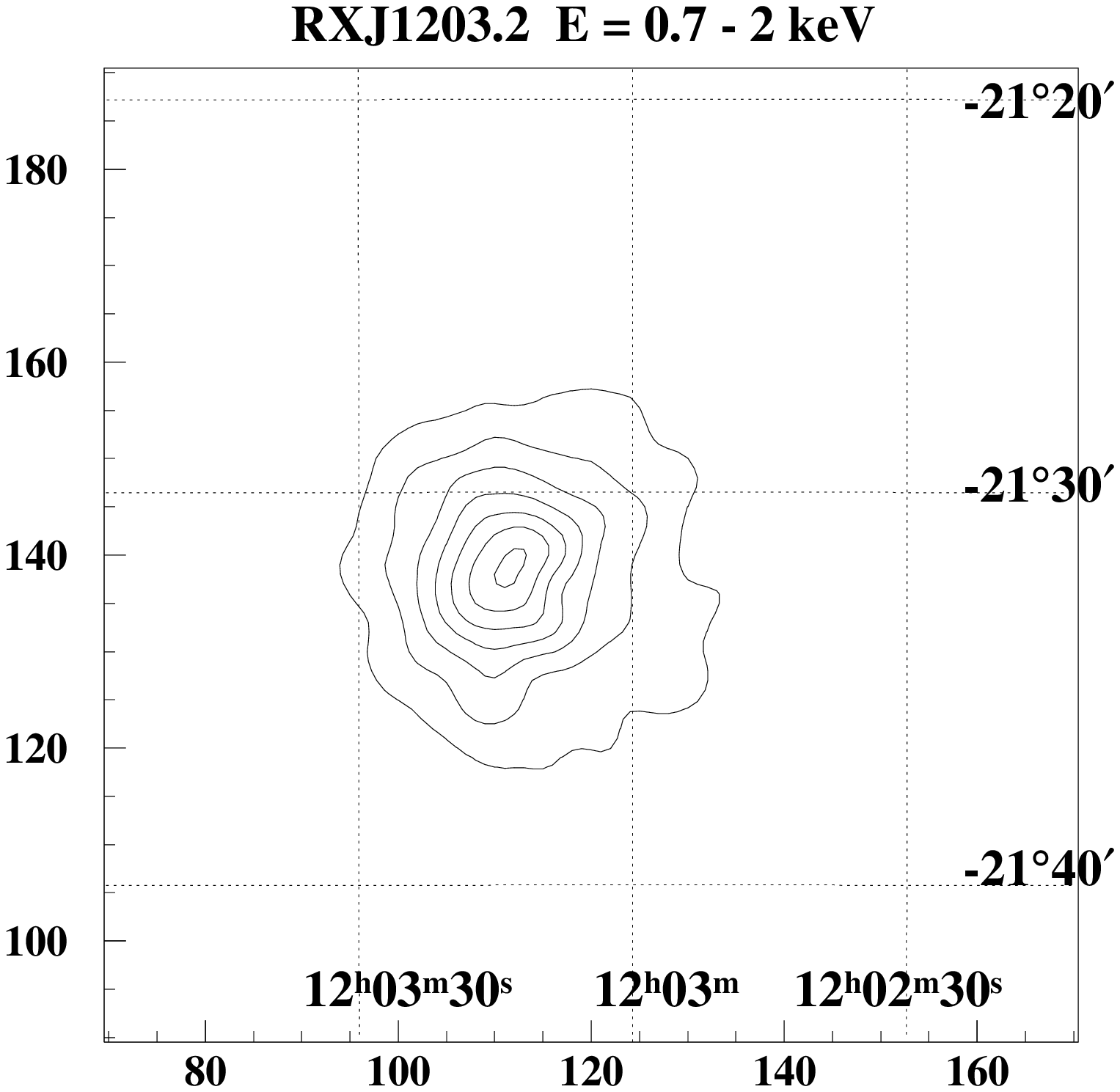}}
\resizebox{.45\hsize}{!}{Fig.1(h)\includegraphics{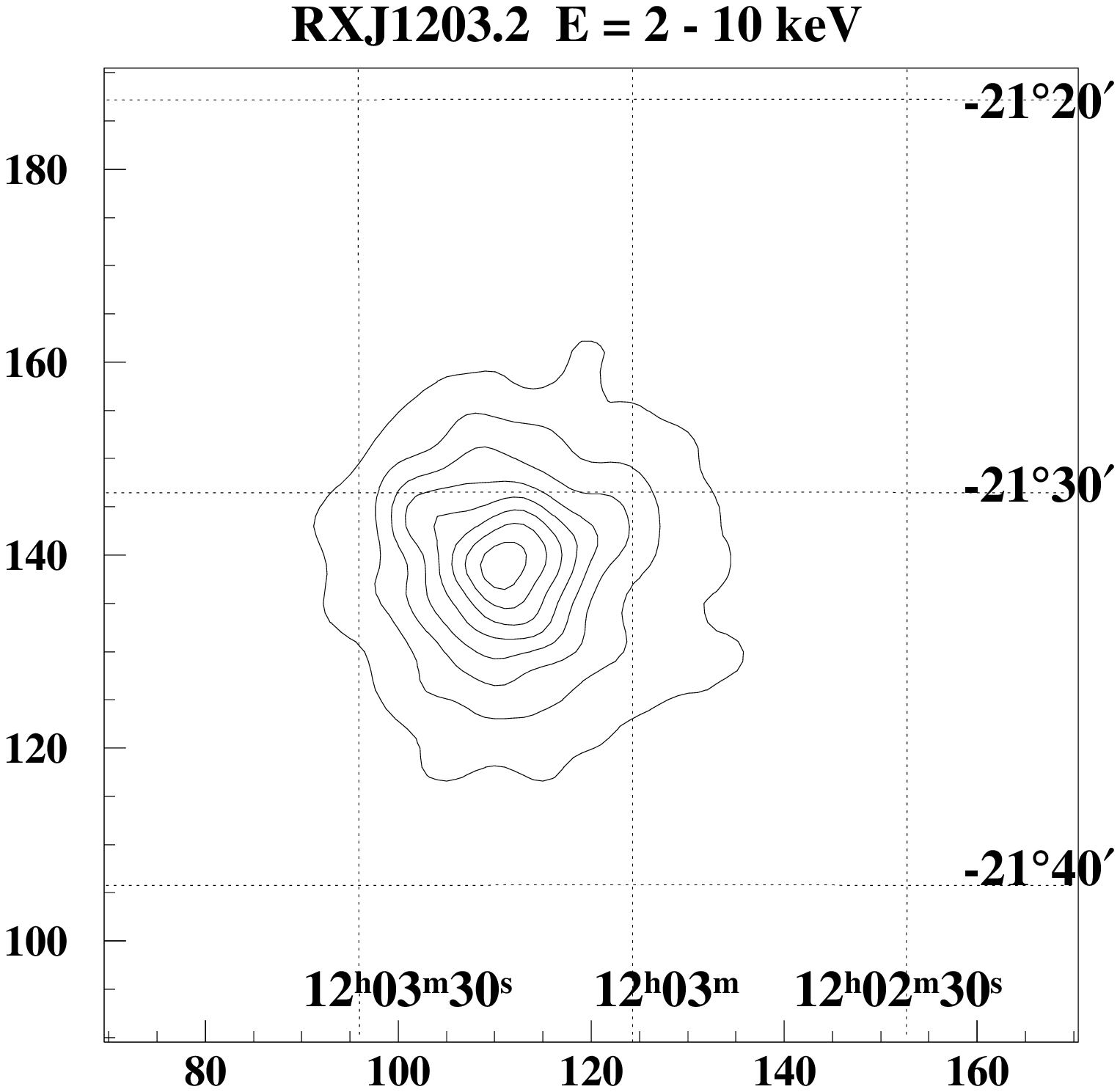}}
\end{figure}

\clearpage

%Fig.2

\begin{figure}
\rotatebox{-90}{
\resizebox{.45\hsize}{!}{Fig.2(a)\includegraphics{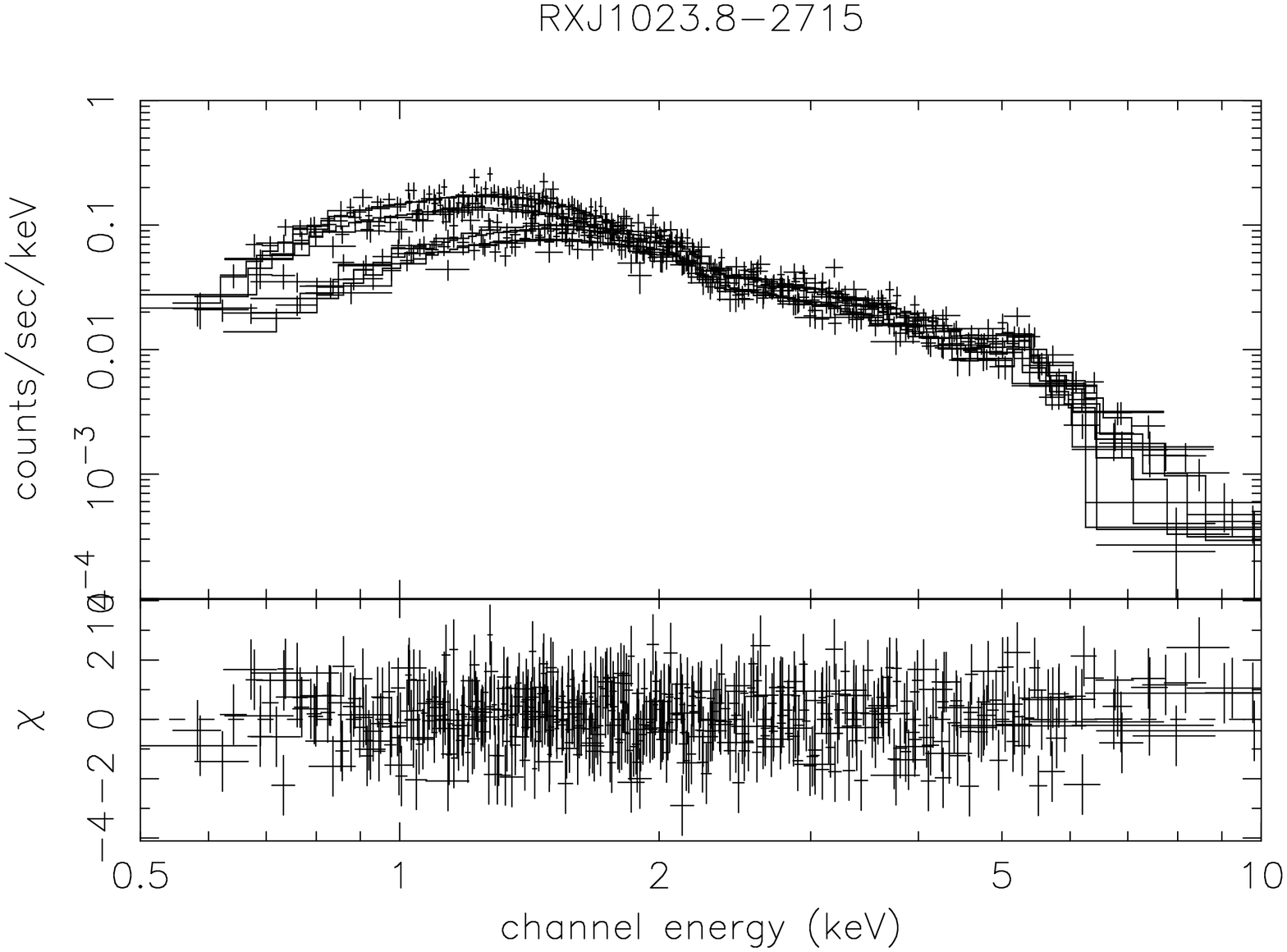}}
}
\rotatebox{-90}{
\resizebox{.45\hsize}{!}{Fig.2(b)\includegraphics{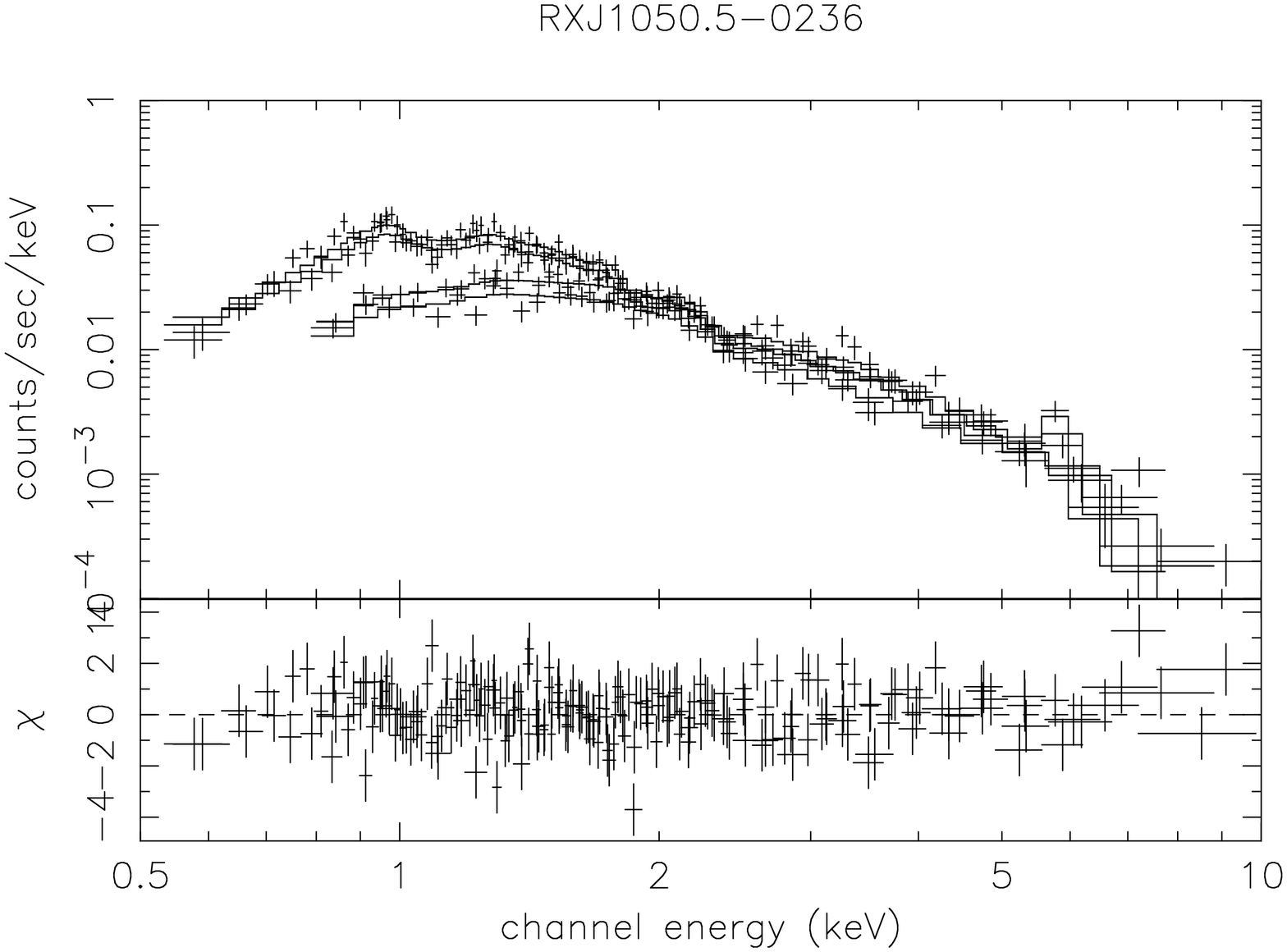}}
}
\end{figure}

%Fig.3

%fig 4

%fig 5

%fig 6

\clearpage

\appendix

\section{Contaminating source for RX~J1050.5-0236}

\begin{figure}
\resizebox{.45\hsize}{!}{\includegraphics{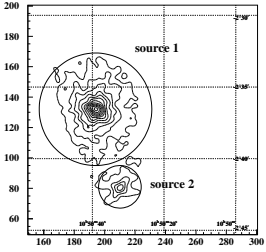}}
\caption{ 
ASCA SIS image of RX~J1050.5-0236 in the 0.6 -- 10.0 keV
band. The pixel size is $6.3''$ and the image has been
filtered by a Gaussian having a $\sigma$ of 1.5
pixel. Contours levels are 1.0, 2.0,
3.0.... counts/pixel. The regions extracted for the spectrum
are indicated by circles.  }
\label{fig:RXJ1050_img}
\end{figure}

\begin{figure}
\rotatebox{-90}{
\resizebox{.45\hsize}{!}{\includegraphics{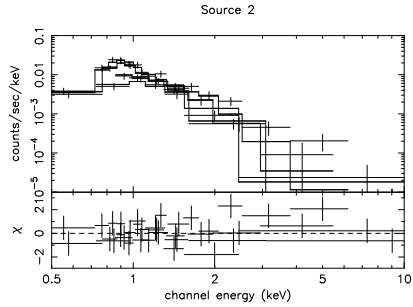}}
}
\caption{ 
ASCA spectra of source 2 in figure \ref{fig:RXJ1050_img}}
\label{fig:RXJ1050s2_spec}
\end{figure}

Fig.~\ref{fig:RXJ1050_img} shows the ASCA SIS (SIS0 + SIS1)
image of RX~J1050.5. The peak designated as ``source 1'' is
RX~J1050.5. Another peak (``source 2'') is conspicuous $\sim
6'$ away from source 1. On the ROSAT HRI image, source 2 is
pointlike and its position ( $(\alpha, \delta)_{\rm J2000}$
= (10h50m27.72s, -2d41m43.1s)) coincides with
\object{HD~93917}, a K0 star (L\'emonon
\cite{Lemonon1999}). Despite the low number of photons in
the ASCA image, we attempt to fit a single-thermal plasma
model to this source: best-fit parameters are $N_{\rm H} =
1.2\err{+1.1}{-0.7}\E{21}$ cm$^{-2}$, $kT =
0.69\err{+0.16}{-0.05}$ keV, $A_{\rm Fe} =
0.14\err{+0.05}{-0.04}$ solar, and $z = 0.0 (<0.057)$.  The
flux in the 0.5 -- 10 keV band is 4.7\E{-13}
erg~s$^{-1}$~cm$^{-2}$. Spectra are displayed on
Fig.~\ref{fig:RXJ1050s2_spec}.

\end{document}